\documentclass[sigconf]{acmart}
\usepackage{array}
\usepackage{booktabs}
\usepackage{makecell}
\usepackage{enumerate}
\usepackage{subcaption}
\usepackage{mwe}
\usepackage[section]{placeins}
\usepackage{CJK}
\usepackage{indentfirst}
\usepackage{amsmath}
\usepackage{multirow}

\usepackage{listings, xcolor}

\definecolor{verylightgray}{rgb}{.97,.97,.97}

\lstdefinelanguage{Solidity}{
	keywords=[1]{anonymous, assembly, assert, balance, break, call, callcode, case, catch, class, constant, continue, constructor, contract, debugger, default, delegatecall, delete, do, else, emit, event, experimental, export, external, false, finally, for, function, gas, if, implements, import, in, indexed, instanceof, interface, internal, is, length, library, log0, log1, log2, log3, log4, memory, modifier, new, payable, pragma, private, protected, public, pure, push, require, return, returns, revert, selfdestruct, send, solidity, storage, struct, suicide, super, switch, then, this, throw, transfer, true, try, typeof, using, value, view, while, with, addmod, ecrecover, keccak256, mulmod, ripemd160, sha256, sha3}, 
	keywordstyle=[1]\color{blue}\bfseries,
	keywords=[2]{address, bool, byte, bytes, bytes1, bytes2, bytes3, bytes4, bytes5, bytes6, bytes7, bytes8, bytes9, bytes10, bytes11, bytes12, bytes13, bytes14, bytes15, bytes16, bytes17, bytes18, bytes19, bytes20, bytes21, bytes22, bytes23, bytes24, bytes25, bytes26, bytes27, bytes28, bytes29, bytes30, bytes31, bytes32, enum, int, int8, int16, int24, int32, int40, int48, int56, int64, int72, int80, int88, int96, int104, int112, int120, int128, int136, int144, int152, int160, int168, int176, int184, int192, int200, int208, int216, int224, int232, int240, int248, int256, mapping, string, uint, uint8, uint16, uint24, uint32, uint40, uint48, uint56, uint64, uint72, uint80, uint88, uint96, uint104, uint112, uint120, uint128, uint136, uint144, uint152, uint160, uint168, uint176, uint184, uint192, uint200, uint208, uint216, uint224, uint232, uint240, uint248, uint256, var, void, ether, finney, szabo, wei, days, hours, minutes, seconds, weeks, years},	
	keywordstyle=[2]\color{teal}\bfseries,
	keywords=[3]{block, blockhash, coinbase, difficulty, gaslimit, number, timestamp, msg, data, gas, sender, sig, value, now, tx, gasprice, origin},	
	keywordstyle=[3]\color{violet}\bfseries,
	identifierstyle=\color{black},
	sensitive=false,
	comment=[l]{//},
	morecomment=[s]{/*}{*/},
	commentstyle=\color{gray}\ttfamily,
	stringstyle=\color{red}\ttfamily,
	morestring=[b]',
	morestring=[b]"
}

\acmConference[ICPC 2022]{The 30th International Conference on Program Comprehension}{May 21–22, 2022}{Pittsburgh, PA, USA}

\AtBeginDocument{%
  \providecommand\BibTeX{{%
    \normalfont B\kern-0.5em{\scshape i\kern-0.25em b}\kern-0.8em\TeX}}}

\begin{document}

\title{Low-level Comments auto-generation for Solidity Smart Contracts}

\begin{abstract}
As self-executing programs on blockchain platforms, smart contracts can build a trusted environment between multi-parties. However, participants who lack programming knowledge usually have difficulties understanding smart contracts by just reading the source code. It brings them difficulties and risks when interacting with smart contracts. A feasible solution is to translate the smart contract source code into natural language descriptions as additional in-line comments to help people better understand, learn and operate smart contracts. This paper proposes an automated translation scheme for Solidity smart contracts, termed SolcTrans, based on an abstract syntax tree and formal grammar. We have investigated 3,000 smart contracts and determined the parts of speech of corresponding blockchain terms. Among them, we further filtered out code snippets without detailed comments and left 811 unique snippets to evaluate the translation quality of SolcTrans. Experimental and user study results show that SolcTrans can accurately translate Solidity codes into comprehensible English texts, which help volunteers to understand smart contracts. 
\end{abstract}


\author{Chaochen Shi}
\affiliation{%
  \institution{Deakin University}
  \country{Australia}}
\email{shicha@deakin.edu.au}



\maketitle

\section{Introduction}

Solidity is a static programming language that runs on the Ethereum Virtual Machine. With Solidity, developers can write self-executing smart contracts and deploy them on Ethereum to build decentralized applications. With the explosive development of decentralized finance (DeFi), the average number of smart contracts deployed each month exceeded 4,200 \cite{duneanalytics} from July 2020 to April 2021. According to the statistics of CoinMarketCap \cite{coinmarketcap}, the DeFi crypto market cap reached 158.5 billion US dollars on 5th January 2022. This promising market immediately attracted the attention of a large number of new users. The number of Ethereum network addresses holding coins is accelerating, and the monthly average number of active Ethereum addresses has exceeded 8.2 million~\cite{duneanalytics}. The development of the DeFi market and the influx of new users have brought considerable concerns, especially for users who are not familiar with smart contracts.

One of the concerns is the technical barrier. Unlike traditional GUI  (Graphic User Interface) software, decentralized applications require users to interact with smart contracts through crypto wallets. Users need to authorize, transfer tokens or provide valid inputs to trigger a smart contract. Users who have difficulty in reading the contract source code may cause misoperation and asset loss. Another concern is false advertising. For example, the Initial Coin Offer (ICO) scams claim to have innovative technology or promising business logic that can provide high returns on investment; however, the smart contracts they actually deployed do not match the content described in their white papers, homepages, or update announcements. As of December 2019, the cumulative losses due to ICO scams have reached 10.12 billion US dollars~\cite{icoscam}. Since most digital assets are minted, issued, and managed by smart contracts, reading the corresponding source code can help users find such inconsistency risks. For most people who lack Solidity programming knowledge, a feasible solution to these two concerns is to translate the Solidity source code into the corresponding natural language description via automated tools. In this way, even users with no programming background can understand and learn the logic details of translated smart contracts, which helps to participate in decentralized applications easily and safely.

There have been some studies about automatic comments generation for Java~\cite{DBLP:journals/tse/McBurneyM16, DBLP:conf/kbse/SridharaHMPV10, DBLP:conf/iwpc/LeClairHWM20} and Python~\cite{DBLP:conf/acl/AhmadCRC20}. However, such deep learning-based approaches can not be directly used in Solidity to solve the concerns mentioned above. First, these approaches generate high-level summaries of the code, which are of limited help in understanding and learning the logical details of the code. Second, deep neural models need high-quality data sets for training, while Solidity has no such public corpus for supervised learning in translation tasks so far.

This paper proposes SolcTrans, an approach focusing on automatically translating Solidity source code into low-level English description. The generated descriptions can be used as in-line code comments to help users understand and learn smart contracts. Since deep learning approaches perform poorly in the current lack of parallel corpus of smart contracts, SolcTrans adopts abstract syntax tree (AST) traversal and context-free grammar (CFG) rules. There are two main challenges for SolcTrans: 

\textbf{How to understand Solidity code through AST?} The AST is a complete tree representation of the hierarchical structure of source code. Each node of an AST corresponds to one or more structures in the source code. Thus we can obtain critical information and logical relationships by traversing the AST, which is more efficient than parsing the original code text. To understand AST nodes, we build an AST translator with a set of customized lexicalization rules. Based on it, AST translator translates different types/levels of AST nodes into corresponding words and phrases, introduced in Section III-D. We also summarize the blockchain terms and specific expressions involved in Solidity and give the corresponding translation templates. 

\textbf{How to generate comprehensible English sentences as the translation outputs?} Given separate words and phrases translated from AST nodes, we need a way to aggregate them into sentences properly as the translation outputs. CFG handles this challenge by constructing hierarchical grammar rules, which are widely used in context-free machine translation tasks~\cite{chan2007word,nguyen2008tree,shen2010string} and also taken by SolcTrans since each Solidity code snippet can be translated independently. We manually marked parts of speech (POSs) of translated AST nodes based on blockchain contexts, then constructed corresponding CFG rules. The final translation results are generated from these CFG rules, comprehensible with proper grammar, and easy to modify.

The contributions of this paper are shown below.
\begin{itemize}
\item We propose a novel machine translation approach, SolcTrans, for translating the Solidity source code into  comprehensible English description. Experimental and user study results show the generated descriptions are accurate and can effectively help users understand the low-level logic of smart contracts.
\item We conduct an extensive study of the Solidity AST, reveal core attributes that can be used for translation, and propose the corresponding translation templates.
\item We collect 3,000 open-source contracts and mark the POSs. The washed data are used to evaluate the performance of SolcTrans. We will release our corpus in the future to facilitate the research in smart contract machine translation.
\end{itemize}

\section{Related work}

To the best of our knowledge, there is no other research work has explored machine translation of Solidity smart contract. In this section, we summarize existing efforts for the similar purposes of this paper, including automatic comment generation and program understanding techniques of the Solidity source code.

\subsection{Automatic comment generation for code}
Some researchers have explored automatic comment generation techniques in Java using the software word usage model (SWUM)~\cite{DBLP:journals/tse/McBurneyM16, DBLP:conf/kbse/SridharaHMPV10}. However, a universal SWUM may struggle to parse Solidity because Solidity is a language designed specifically for blockchain operations, which involve many reserved terms. Other state-of-the-art solutions are to use deep~\cite{DBLP:conf/iwpc/HuLXLJ18, DBLP:conf/icml/AllamanisPS16, DBLP:conf/acl/AhmadCRC20} or graph~\cite{DBLP:conf/iwpc/LeClairHWM20} neural networks to automatically generate high-level comments. The encoder-decoder structure is widely used in related research. The encoder encodes the source code into a vector with a fixed size, while the decoder predicts comments based on vectors. The internal structure of codecs could be a CNN and variants of an RNN, such as the gated recurrent unit (GRU) \cite{DBLP:journals/corr/abs-1709-07642} and long short-term memory (LSTM) model \cite{DBLP:conf/acl/IyerKCZ16}. Nahla et al.~\cite{DBLP:conf/icsm/AbidDCM15} proposed an approach to generate natural-language summary for C++. It uses predefined templates to produce generic summaries and specialize them through static analysis. The generated summaries focus on main components of methods and ignore some logic details.

\subsection{Program comprehension of Solidity}
The current research on the program comprehension of Solidity is focused mainly on static analysis, which is widely used in vulnerability detection. The mainstream smart contract vulnerability detection tools include Oyente \cite{DBLP:conf/ccs/LuuCOSH16}, Zeus \cite{DBLP:conf/ndss/KalraGDS18}, and SmartCheck~\cite{DBLP:conf/icse/TikhomirovVITMA18}. Oyente uses symbolic execution to establish a control flow graph and statically analyze the smart contract based on the graph; Zeus proposes a formal verification framework based on which users can verify the correctness and fairness of smart contracts. SmartCheck~\cite{DBLP:conf/icse/TikhomirovVITMA18} is another static analysis tool for Ethereum smart contracts. SmartCheck performs grammar and lexical analysis on the source code and uses the Extensible Markup Language (XML) to describe the AST. It uses the XML Path Language (Xpath) to discover vulnerabilities and other safety issues.

Moreover, there are (deep) neural network approaches about making smart contracts more understandable. ScanAT~\cite{DBLP:journals/access/KimPL19} infers the pre-defined attribute tags of a smart contract from its bytecode through the trained neural network. These attribute tags help people to understand the actual functions implemented in a smart contract. Chen et al.~\cite{DBLP:conf/www/ChenZCNZZ18} proposed an model to detect whether a smart contract is a Ponzi scheme based on its account features and code features. Unlike these studies, our approach focuses on machine translation for Solidity to make people better understand smart contracts at the source code level.
\section{Proposed Approach}
This section describes the translation scheme of proposed SolcTrans and the details of each module.
\subsection{Overview}
\begin{figure}
  \centering
  \includegraphics[width=0.5\textwidth]{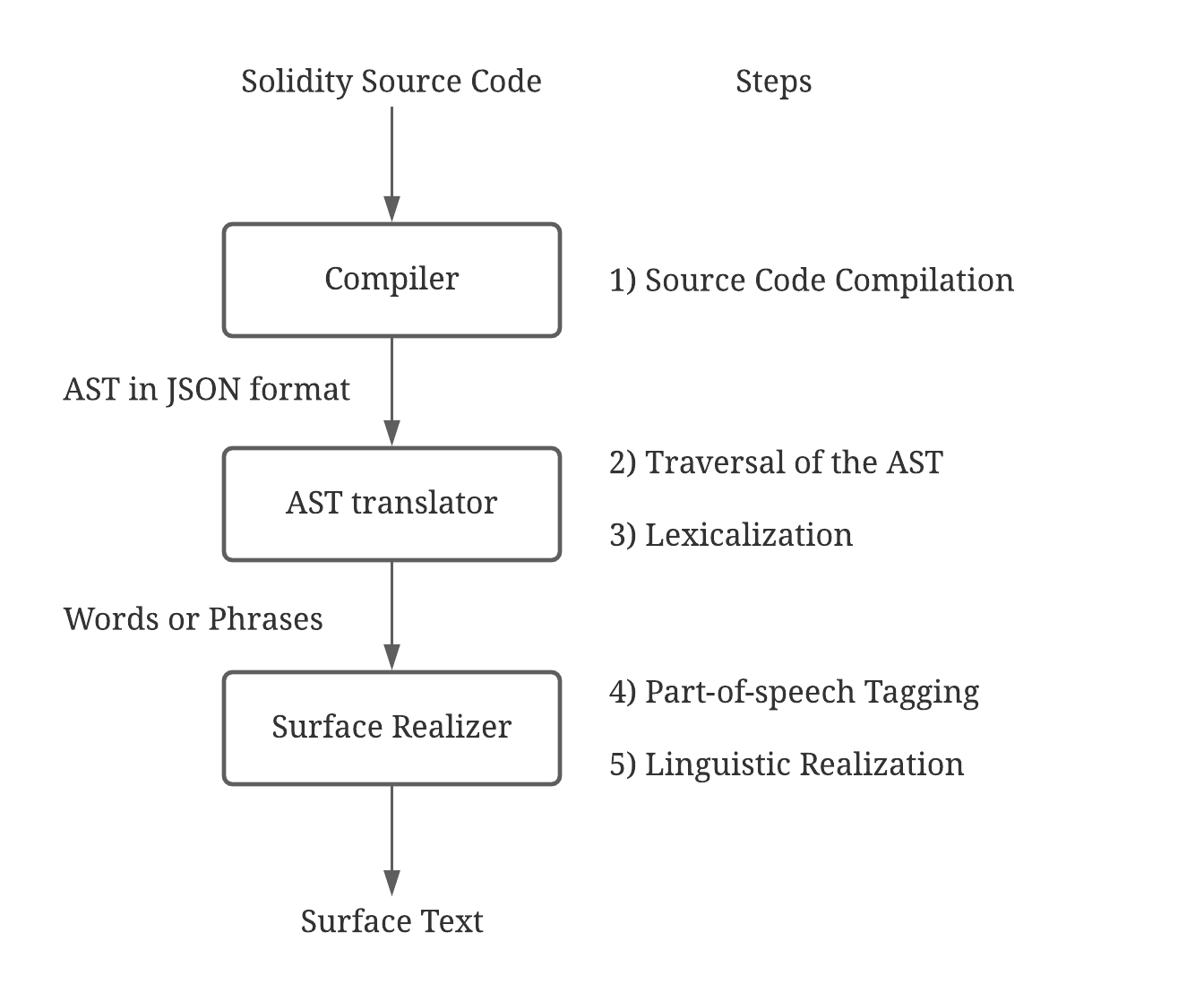}
  \caption{The translation scheme of SolcTrans.}
  \label{fig:archi}
\end{figure}

The translation scheme of SolcTrans is illustrated in Fig.~\ref{fig:archi}. There are two core modules: the AST translator and the surface realizer. We first collect verified smart contract source code by crawling from Ethereum block explorers and Google BigQuery. Then, we compile and analyze the collected smart contracts to understand the AST output of Solidity and to determine which word will be used in the translations. We build our AST translator in accordance with these patterns. Once we obtain the phrases generated by the AST translator, we aggregate them into readable sentences, known as ``surface text''. Specifically, the surface realizer annotates the POSs of the phrases and tries to generate the surface texts based on predefined grammar rules. 

The workflow of SolcTrans is summarized into five main steps as Fig.~\ref{fig:archi} shows. First, we use the official Solidity compiler \emph{solc} to generate the AST of a smart contract and dump it into the JSON format. The second and third steps are carried out by the AST translator, including traversing the AST and translating AST nodes into English words or phrases. Then these words or phrases are tagged with the corresponding POSs in a standard format. Finally, the surface realizer aggregates these words properly and generates the surface texts.

\subsection{Obtaining AST and core attributes}
\begin{figure}
  \centering
  \includegraphics[width=0.5\textwidth]{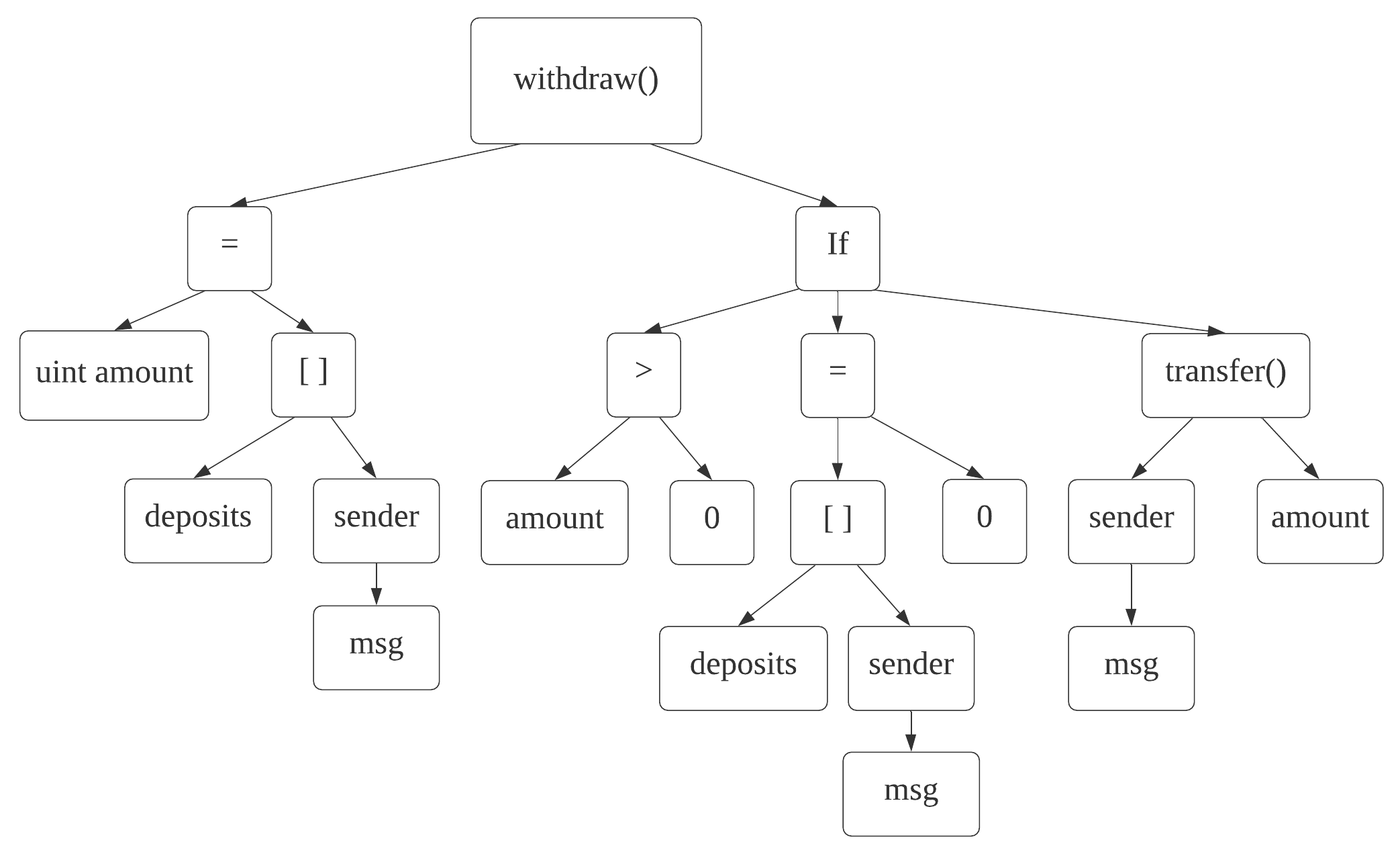}
  \caption{The AST diagram of Listing 1.}
  \label{fig:ast}
\end{figure}

The entire AST is composed of multiple nested nodes, while each type of node has different properties, describing Solidity’s equivalents of elements. Listing 1 shows a function that allows users to withdraw their deposits from the smart contract. The function contains statements such as assignment, condition, comparison, and function call. The corresponding AST hides these logical expressions in nested structures, as Fig.~\ref{fig:ast} shows. The official Solidity compiler \emph{solc} provides a function to export the AST of the source code in JSON format using the command ``\verb|--ast-compact-json|''. Listing 2 shows the detailed information of the node \emph{uint amount} in Fig.~\ref{fig:ast} as an example. 

\begin{lstlisting}[caption = A withdraw function in Solidity][language=Solidity]
function withdraw() public {
    uint amount = deposits[msg.sender];
    if (amount > 0) {
        deposits[msg.sender] = 0;
        msg.sender.transfer(amount);
    }
}
\end{lstlisting}

\lstset{
    string=[s]{"}{"},
    stringstyle=\color{blue},
    comment=[l]{:},
    commentstyle=\color{black},
}
\begin{lstlisting}[caption = The JSON format of the node \emph{uint amount} of Listing 1]
{
	"constant" : false,
	"id" : 9,
	"name" : "amount",
	"nodeType" : "VariableDeclaration",
	"scope" : 36,
	"src" : "137:11:0",
	"stateVariable" : false,
	"storageLocation" : "default",
	"typeDescriptions" : {...},
	"typeName" : {...},
	"value" : null,
	"visibility" : "internal"
}
\end{lstlisting}
Each AST node contains an attribute \emph{nodeType} that defines the purpose of the node. According to the value of \emph{nodeType}, we identified the core attributes which are used for code translation. For example, the \emph{nodeType} of the field \emph{uint amount} shown in Listing 2 is \emph{VariableDeclaration}, while the core attributes are \emph{name} and \emph{typeDescriptions}. Similarly, the \emph{nodeType} of the method is \emph{FunctionDefinition}, and the core attributes are self-explanatory, i.e., \emph{body}, \emph{parameters}, and \emph{returnParameters}. In addition, there are other common node types, which we summarized in Table~\ref{tab:nodeTypes}, shown in Appendix A. It is worth noting that some attributes with values of nested structures are child nodes of the current node, such as \emph{typeDescriptions} in Listing 2 and \emph{body} in a function node. 

\subsection{Node preprocessing}
The original AST contains nodes corresponding to the version numbers and compilation information. These nodes are meaningless for code translation; thus, we regard them as redundant nodes. To improve the efficiency of code translation, we need to preprocess the AST to exclude redundant nodes and keep only nodes that are closely related to the translation. The specific steps of preprocessing are as follows:

\begin{enumerate}[(1)]
\item SolcTrans recognizes a node as the root node if the value of \emph{nodeType} is \emph{SourceUnit}. The root node contains all the contract information; thus, SolcTrans keeps it and traverses its child nodes.
\item SolcTrans excludes the child nodes of type \emph{ImportDirective} and \emph{PragmaDirective}, which represent the imported dependencies and the version of the compiled code, respectively. SolcTrans retains only the nodes of type \emph{ContractDefinition} since they contain the main body of the contract.
\end{enumerate}
After preprocessing, we obtained nonredundant nodes related to code translation. SolcTrans further traverses these reserved nodes for subsequent translation.

\subsection{AST traversal}
Once we obtain the preprocessed AST, the AST translator of SolcTrans performs a depth-first traversal of the reserved nodes to translate the nodes into English words or phrases. The AST translator can directly translate leaf nodes into words or fixed expressions as outputs. By combining leaf node outputs in specific patterns, the AST translator can also translate the parent nodes or grandparent nodes of the leaf nodes into phrases. For a node with a higher level, the output is a list containing multiple phrases, while the elements in the list correspond to the translation results of lower-level nodes.

\begin{figure}[t]
  \centering
  \includegraphics[width=0.5\textwidth]{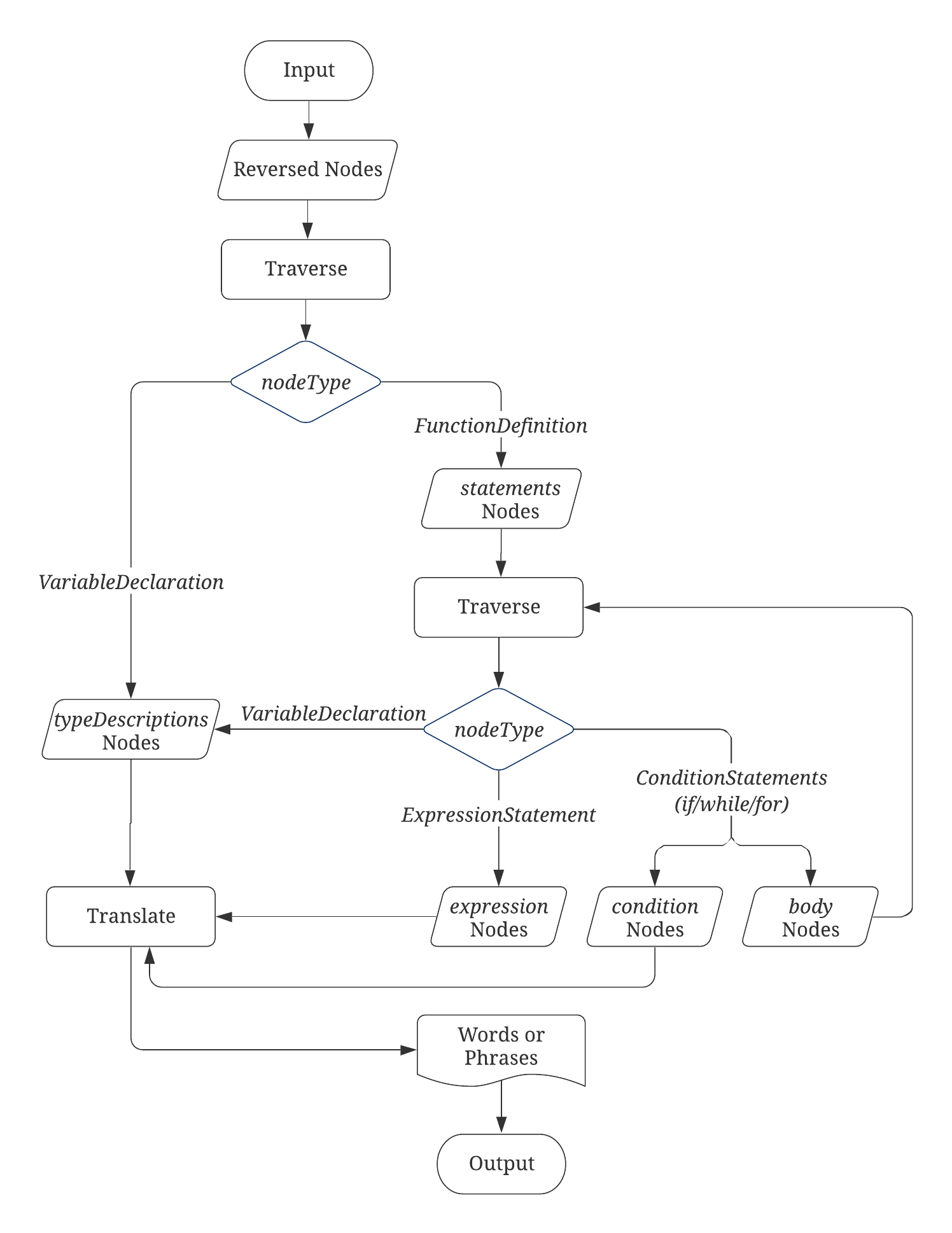}
  \caption{The process of AST traversal.}
  \label{fig:traverse}
\end{figure}

Fig.~\ref{fig:traverse} shows the process of AST traversal. More specifically, the AST is traversed following the steps below:
\begin{enumerate}[(1)]
\item Traverse the nodes of the preprocessed AST with the deep-first procedure, and check the value of the \emph{nodeType} of each node. If the \emph{nodeType} is \emph{VariableDeclaration}, the node represents a variable declaration statement. SolcTrans can directly translate its child node \emph{typeDescriptions} which contains the type information of the variable. If the \emph{nodeType} is \emph{FunctionDefinition}, then the node is the root of a function. Its child nodes \emph{statements} represent structures in the function, which require further parsing.
\item Traverse the \emph{statements} nodes and check their \emph{nodeType}. If the \emph{nodeType} is \emph{ExpressionStatement}, the node represents an expression of the source code and will be parsed in Step 3. If the \emph{nodeType} is \emph{ConditionStatements}, the node represents a control structure that contains a \emph{for}/\emph{while}/\emph{if} statement. Its child nodes \emph{condition} can be directly translated, while the nodes \emph{body} should be parsed via Step 2 again.
\item The node with type \emph{ExpressionStatement} has the only child node \emph{expression} which contains all the elements of the corresponding expression. SolcTrans translates the node \emph{expression} into phrases with the predefined pattern corresponding to its \emph{nodeType}.
\end{enumerate}

The predefined translation patterns of low-level nodes are as follows:\\
\\
{\bfseries \emph{typeDescriptions} nodes:} As mentioned in Step 1, the \emph{typeDescriptions} node contains the type information of the variable declared in its parent. SolcTrans gets attribute \emph{typeString} from the \emph{typeDescriptions} node and gets attribute \emph{name} from its parent. A variable declaration statement is directly translated as ``Variable \emph{name} is declared as a/an \emph{typeString}''. For example, \textsf{uint count = 0} is translated as ``Variable count is declared as an uint''. \\
\\
{\bfseries \emph{condition} nodes:} The \emph{condition} node represents a conditional expression of \emph{if}, \emph{while}, \emph{doWhile}, and \emph{for} structures. Usually, it is a judgment that returns a Boolean value. SolcTrans gets attributes from the \emph{condition} node and links them with nested structures and according to their \emph{nodeTypes}. For example, the node corresponding to the statement ``\textsf{for (i = 0; i < N; i++) \{\ldots\}}'' has four attributes: \emph{initializationExpression} (\textsf{i = 0}), \emph{condition} (\textsf{i < N}), \emph{loopExpression} (\textsf{i++}), and \emph{body} (\textsf{\ldots}). The translation template would be ``Set \emph{condition}, then as long as \emph{condition}, \emph{body}. Each time that happens \emph{loopExpression}.'' A qualitative example can be found in Table~\ref{tab:examples}.\\
\\
{\bfseries \emph{expression} nodes}: SolcTrans checks the value of \emph{nodeType} and translates the \emph{expression} node in specific patterns. We list several common \emph{nodeTypes} of \emph{expression} nodes as examples.
\begin{enumerate}[-]
\item \emph{Identifier}: The node represents a global or local variable; thus, SolcTrans directly gets and returns the value of the attribute \emph{name}. If the value is named in the camel case or snake case, SolcTrans parses it into words based on POS. For example, the variable \emph{messageSender} is translated into two nouns: \emph{message} and \emph{sender}.
\item \emph{Assignment}: The node represents an assignment statement, which contains the core attribute \emph{leftHandSide}, \emph{rightHandSide} and \emph{operator}. SolcTrans treats \emph{leftHandSide} and \emph{rightHandSide} as the \emph{expression} nodes and describes the assignment operator with fixed templates.
\item \emph{FunctionCall}: The node represents calling a function. SolcTrans gets function names from nested child nodes \emph{MemberAccess} and gets arguments from the attribute \emph{arguments}. For a function \emph{func(args)}, the translation is ``(call/result of) the function \emph{func} with arguments \emph{args}'' depending on the context.  

\end{enumerate}
The parse procedures of most unlisted nodes are similar to the nodes in Fig. 3. The only difference is the depth of traversal and the core attributes. For example, the keyword \emph{modifier} indicates a behavior to change a function. SolcTrans explains the \emph{modifier} with predefined descriptions and traverses its child nodes as a way of traversing the body of a function.

\subsection{Special statements translation}
There are some special statements in Solidity that are closely related to blockchain features. These statements, including global variables and system functions, have specific meanings, but it is difficult to understand them by their names directly. Thus, we designed customized templates for translating these variables and functions mentioned in Solidity Docs V0.8.4~\cite{Sol}.

For example, the global variable \emph{msg} contains attributes that allow access to the blockchain. Its attribute \emph{msg.sender} refers to the address of the external function call; \emph{msg.value} refers to the amount of ether sent to the contract. Thus \emph{msg.sender} and \emph{msg.value} are translated as ``user'' and ``the money sent by the user'', respectively. Similarly, the global variable \emph{block} contains attributes of the current block. Its representative attributes \emph{block.difficulty} and  \emph{block.gaslimit} are translated as ``the current difficulty for mining a block'' and ``the limit of the total gas usage in the current block'', respectively. Another example is the system function \emph{gasleft() returns(uint256)} which 
is translated as ``the remaining gas of the current call'' to help readers understand the gas related operation. More examples of special statements translation are shown in Table~\ref{tab:sps}, shown in Appendix B.

\subsection{Natural language generation}
Once we obtained the phrases and words output by the AST translator, we need to aggregate them into comprehensible sentences according to the correct grammar as surface texts. Specifically, the surface realizer of SolcTrans performs POS tagging on the obtained words and phrases and then generates sentences based on the context-free grammar (CFG) we constructed.

The surface realizer identifies the POS of each word and phrase, including words decomposed from camel case and snake case phrases. We tag each POS in the standard format of the Penn Treebank Tagset~\cite{taylor2003penn}. For words with multiple POSs, we determine the POS of each according to the context. For example, the variable name \emph{highestBid} is parsed into two words \emph{highest} and \emph{bid} by the AST translator. Among them, the word \emph{highest} is marked as $JJS$, which indicates a superlative adjective. The POS tag of the word \emph{bid} could be a verb or noun. Here, we predict the word \emph{bid} as $NN$ (a singular noun) because an adjective modifies it. In addition, the corpus of the Penn Treebank Tagset comes from the statistical data of the Wall Street Journal and lacks data related to blockchain terms. We manually add the POSs of blockchain terms and abbreviations so that the Penn Treebank Tagset can cover the words that appear in Solidity. For example, the abbreviation ``tx'' for the word ``transaction'' is marked as $NN$.

We construct a CFG to generate sentences based on the Penn Treebank tags. Specifically, $R$ is a set of rules, each of which can be represented by mapping a variable $X$ to a string $Y_i$:\\
\begin{equation}
   X\rightarrow Y_1,Y_2,\ldots, Y_i,     X\in N,    Y_i\in (N \cup \sigma)
\end{equation}
\begin{enumerate}[-]
\item $N$ represents a set of POS tags in the Penn Treebank standard format, such as \{$NP, VP, NN, \ldots$\};
\item $\sigma$ represents a set of words defined in the Penn Treebank Tagset;
\end{enumerate}

\begin{figure}
  \centering
  \includegraphics[width=0.5\textwidth]{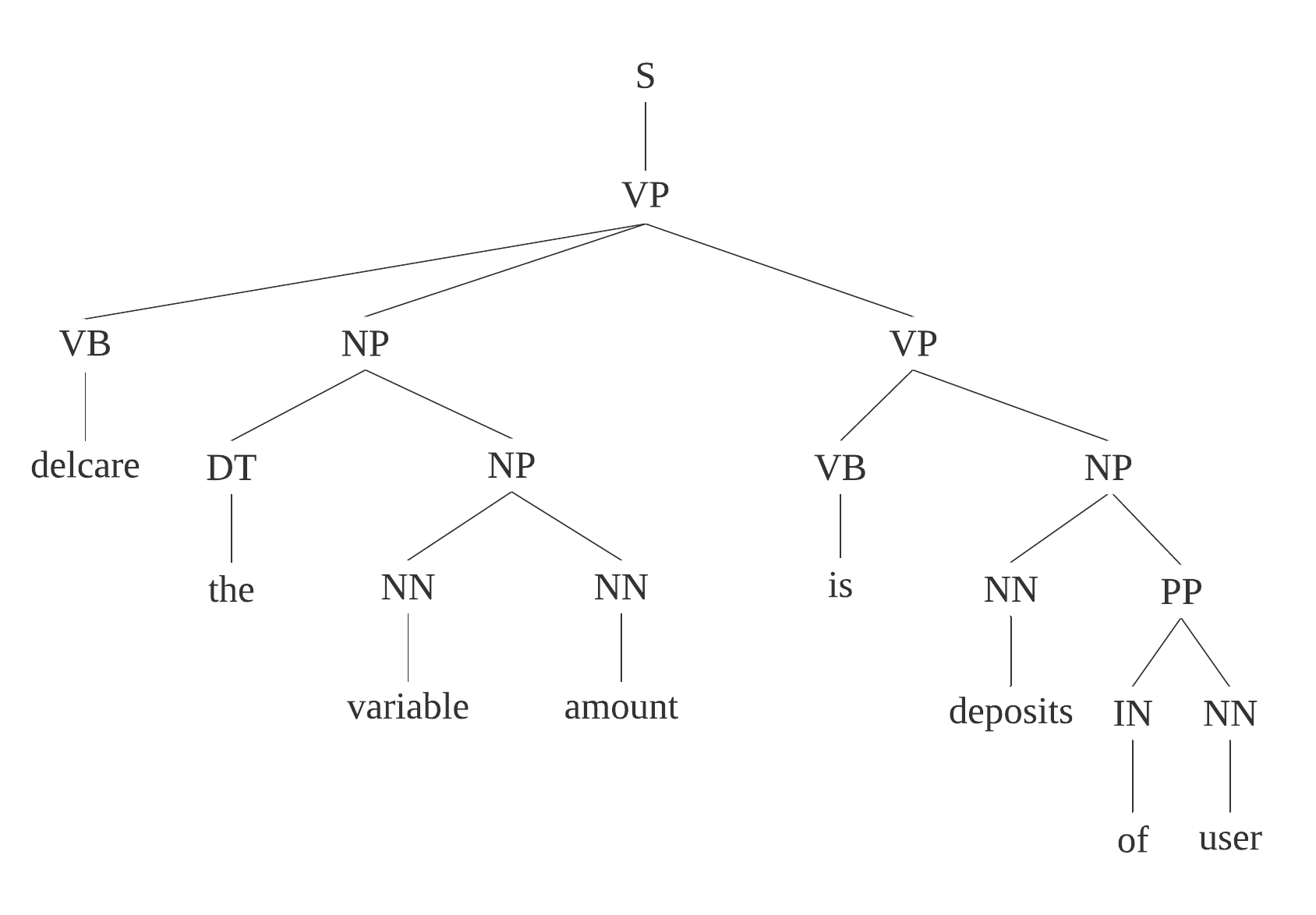}
  \caption{The syntax tree of the example sentence.}
  \label{fig:cfg}
\end{figure}

A simplified example of CFG used to translate Listing 1 is defined as\\
\\
~~$R =~{\left\{
{\begin{aligned}
&S\rightarrow VB \mid NP \mid VP\\
&VP\rightarrow VP~PP \mid VB~IN \mid VB~NP\\
&NP\rightarrow DT~NP \mid DT~NN \mid NN~NN  \mid NN~PP\\
&PP\rightarrow IN~NN
\end{aligned}}
\right.} $
\\

where $N = \{S, NP, VP, PP, DT, VB, NN, IN\}$, and $S$ is a start symbol of a sentence. Once SolcTrans gets a collection of words, it constructs sentences recursively from these words according to the CFG. For example, line 2 of the function in Listing 1 is parsed into a nested list of words \emph{\{\{the variable amount, declare\}, \{is\}, \{deposits of user\}\}} after traversal. Thus, the set of words is \\
\\
$\sigma =~{\left\{{\begin{aligned}
&VB\rightarrow declear \mid is\\
&NP\rightarrow the~variable~amount \mid deposits~of~user\\
&NN\rightarrow variable \mid amount \mid deposits \mid user\\
&DT\rightarrow the\\
&IN\rightarrow to \mid of
\end{aligned}}
\right.} $
\\
\\
Given $R$ and $\sigma$, the surface realizer can reverse-engineer words and phrases to determine which rules in $R$ are used to aggregate these elements of sentences. For example, ``is'' is a $VB$ (verb, base form), and ``deposits of user'' is an $NP$; thus, they can be combined with a $DT$ ``the'', producing another $VP$ (verb phrase) ``is deposits of user''. The complete sentence generated by $R$ is ``declare the variable amount is deposits of user''. The syntax tree is shown in Fig.~\ref{fig:cfg}.

Using CFG, we can deduce a sentence's grammatical structure or aggregate sentences that describe multiple code lines. Nevertheless, the derived syntax may be ambiguous. Common ambiguity problems are as follows:
\begin{enumerate}
\item Different POSs of words, e.g., the word ``bid'' can be marked as either a verb or noun;
\item The scope of the prepositional phrase, e.g., for the structure of $VP$ $PP$ $PP$, the second prepositional phrase may describe the VP or the first $PP$;
\item Consecutive nouns, such as $NN$ $NN$ $NN$.
\end{enumerate}
To conduct disambiguation of the derived syntax, we need to find the most probable tree from various of possible syntax trees. Discriminative disambiguation algorithms, such as Max-Margin Markov Networks~\cite{DBLP:conf/nips/TaskarGK03} and Conditional Random Field~\cite{DBLP:conf/icml/LaffertyMP01}, require complete feature engineering which is hard to implement in our limited corpus. Thus, we use the generative algorithm --- probabilistic CFG (PCFG) here since it is more robust in a limited labeled data set~\cite{DBLP:reference/ml/Sakakibara17}. Specifically, we assign a probability $p(r)$ to each rule in R based on its frequency. For each possible syntax tree, we take the product $p(t)$ of $p(r)$ in it as its probability of being selected. Thus the selected tree is $arg~max~p(t)$. To compute $p(t)$, we need to perform the following:
\begin{enumerate}
\item Collect all $N$ and $\sigma$ in the corpus;
\item Collect all the rules in the corpus as R;
\item For each rule $X\rightarrow Y$, compute $p(r) = p(X \rightarrow Y) / p(X)$ from the corpus;
\end{enumerate}

The size of the fully commented corpus is too small to support our current PCFG in covering all the syntax. Therefore, we use both CFG and PCFG in most cases and will continue to improve the PCFG by expanding the corpus.

\section{Experiments and analysis}

We conducted experiments to evaluate the performance of SolcTrans. We focused on two questions: 
\begin{itemize}
\item Q1: What is the translation quality of SolcTrans?
\item Q2: Is its translation quality affected by the complexity (length or gas cost) of the code?
\end{itemize}
 All the experiments in this paper are implemented with Python 3.7, and run on a PC with a 2.2 GHz Intel Core i7 CPU, 32 GB 3200MHz DDR4 RAM.
\begin{figure*}
    \centering
    \begin{subfigure}{0.45\linewidth}
        \includegraphics[width=\linewidth]{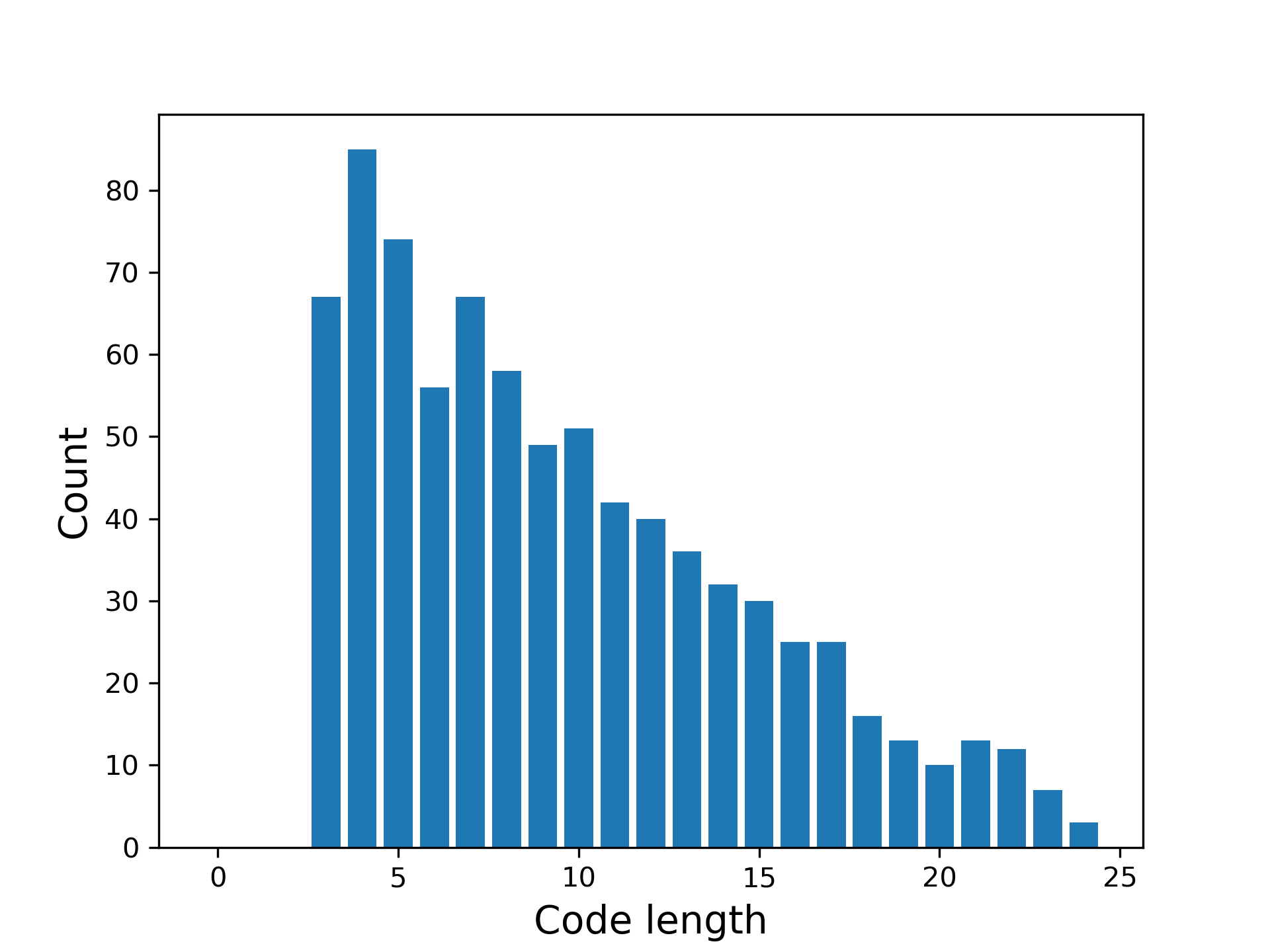}
        \caption{Length distribution of code snippets.}\label{fig:codeCT}
    \end{subfigure}
    \begin{subfigure}{0.45\linewidth}
        \includegraphics[width=\linewidth]{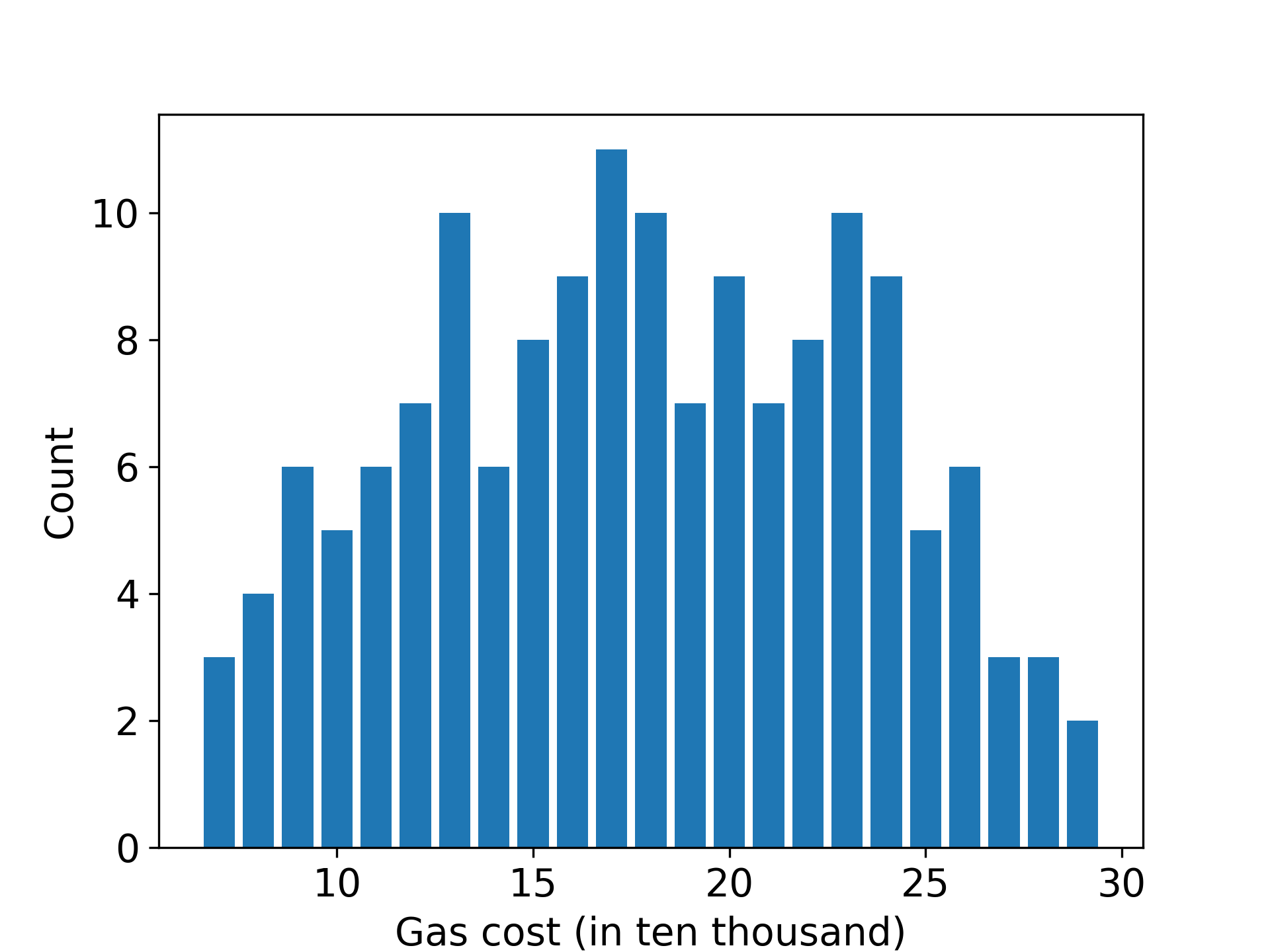}
        \caption{Gas cost distribution of functions.}\label{fig:gasCT}
    \end{subfigure}
    \caption{Distributions under different code lengths and gas cost (counted in gwei).}\label{fig:count}
\end{figure*}

\subsection{Data Preparation}
There is no public Solidity corpus containing code-description pairs so far. To build an experimental data set, we collected 3,000 open-source smart contracts of the top 100 Ethereum Dapps ranked by their user activities (unique source addresses in transactions to Dapp contracts) over the past 30 days as of 1st May 2021. All smart contracts were collected from the Ethereum explore \emph{etherscan.io} through web crawlers. According to the statistics, the vocabulary size of the code and description was 1,272 and 2,140, respectively. We marked the POSs of these words to expand the Penn Tree Tagset for the Solidity code translation task. After that, we removed duplicates from the data set and only kept one copy of them. Code snippets without corresponding descriptions were also excluded. Finally, 811 code snippets and corresponding descriptions (used as reference sentences in experiments) were reserved as the data set for evaluating the performance of SolcTrans. 

The length of a code snippet was measured by the number of its tokens. The camel case and snake case phrases were divided into individual words (tokens), and the special symbol was counted as a token. The length distribution of single code snippets in our corpus is shown in Fig.~\ref{fig:codeCT}. 

The gas cost of a code snippet was estimated by a global function \emph{estimateGas()} defined in web3.js, counted in gwei (1 gwei = $10^{-9}$ ETH). Note that only the functions have estimable gas cost, so we counted the distribution of the gas cost corresponding to 154 complete functions from all the 811 code snippets as shown in Fig.~\ref{fig:gasCT}.

\subsection{Evalution Metrics}
We use three metrics that are widely utilized in the area of machine translation to evaluate the translation quality of SolcTrans: BLEU-N~\cite{DBLP:conf/acl/PapineniRWZ02}, ROUGE-L~\cite{lin2004rouge}, and METEOR~\cite{DBLP:conf/acl/BanerjeeL05}. BLEU-N measures the N-gram precision between candidate sentences and reference sentences; N could be 1, 2, 3, or 4. When N = 1, the metric indicates the accuracy of individual words in the translation; when N \textgreater~1, the metric indicates the similarity between the translation and the reference sentence. ROUGE-L is a metric that matches the longest common sequence between two sentences and returns the recall rate. METEOR considers both accuracy and recall rate and returns the F value. METEOR also supports the matching of synonyms by a given thesaurus. Here, we use the default thesaurus form WordNet~\cite{miller1998wordnet}. These metrics fall between 0 and 1 with a higher number indicating better translation quality. Because the sentences generated by the correct CFG are generally coherent, we are more concerned with whether there are omissions or ambiguities. Therefore, BLEU-1 and ROUGE-L have higher weights among all the metrics. 
\subsection{Experimental Analysis}

Table~\ref{tab:evalution} presents the details of the experimental results to answer Q1. The BLEU-1 score indicates that more than one-third of the words in the translations are included in the reference sentences. It shows the high accuracy of words translated by SolcTrans. The BLEU-2, BLEU-3, and BLEU-4 scores are relatively low. This means that the way SolcTrans constructs sentences and phrases is usually different from the reference sentences, resulting in few overlapping clauses between them. This also affects the value of ROUGE-L, makes it mainly contributed by ROUGE-1. The ROUGE-L and METEOR scores indicate that there are a few omissions in the translation results, and some omissions are actually synonymous expressions of the reference sentence. Some synonyms of blockchain terms are not included in WordNet, which lowers the METEOR score.

\begin{table}[h]\footnotesize
    \caption{Evaluation results of SolcTrans.}
    \label{tab:evalution}
    \begin{tabular}{ccclll}
        \hline
        BLEU-1 & BLEU-2 & BLEU-3 & BLEU-4                     & ROUGE-L                    & METEOR                     \\ \hline
        0.3472 & 0.0946 & 0.0608 & \multicolumn{1}{c}{0.0144} & \multicolumn{1}{c}{0.3784} & \multicolumn{1}{c}{0.0808} \\ \hline
\end{tabular}
\end{table}

\lstset{linewidth=7cm}
\begin{table*}
  \renewcommand\arraystretch{1}
  \caption{Qualitative examples of translation outputs.}
  \label{tab:examples}
  \begin{tabular}{l @{}m{4.6cm}<{\raggedright} m{4.5cm}<{\raggedright}}
    \toprule
    \makecell[c]{Code snippet} & \makecell[c]{Generated sentence} &  \makecell[c]{Reference sentence}\\
    \midrule
    \specialrule{0em}{1pt}{-1pt}
    \begin{lstlisting}[language=Solidity,numbers=none]
uint amount = deposits[msg.sender];
\end{lstlisting} & Declare the variable amount is deposits of user. & Set variable amount as user's deposits.\\\hline

    \begin{lstlisting}[language=Solidity,numbers=none]
if (amount > 0) {
        deposits[msg.sender] = 0;
        msg.sender.transfer(amount);
    }
\end{lstlisting} & If amount is greater than 0, deposits of user is 0 and transfer amount to user. & If amount is greater than 0, set user's deposits to 0 and then transfer all the left amount to user.\\\hline

    \begin{lstlisting}[language=Solidity,numbers=none]
for (uint i = 0; i < proposalNames.length; i++) {
        proposal.push(name[i]);
    }
\end{lstlisting} & Set i is 0, then as long as i is less than the length of proposal names, push the name of i to proposal. Each time that happens add one to i. & When i is less than the length of proposal names, push the name at index i to the end of proposal.\\\hline
    \begin{lstlisting}[language=Solidity,numbers=none]
modifier onlyBuyer() {
        require(msg.sender == buyer);
        _;
    }
\end{lstlisting} & Confirm user equals to buyer before executing onlyBuyer(). & Confirm that buyer is the current call before calling the function.\\\specialrule{0em}{1pt}{1pt}\hline
\begin{lstlisting}[language=Solidity,numbers=none]
return keccak256(abi.encode(a, b, c, d));
\end{lstlisting} & Return Ethereum-SHA-3 (Keccak-256) hash of the a, b, c, d encoded by abi. & Return the keccak256 hash value of encoded a, b, c, d.\\
    \bottomrule
  \end{tabular}
\end{table*}

\begin{figure*}[t]
    \centering
    \begin{subfigure}{0.45\linewidth}
        \includegraphics[width=\linewidth]{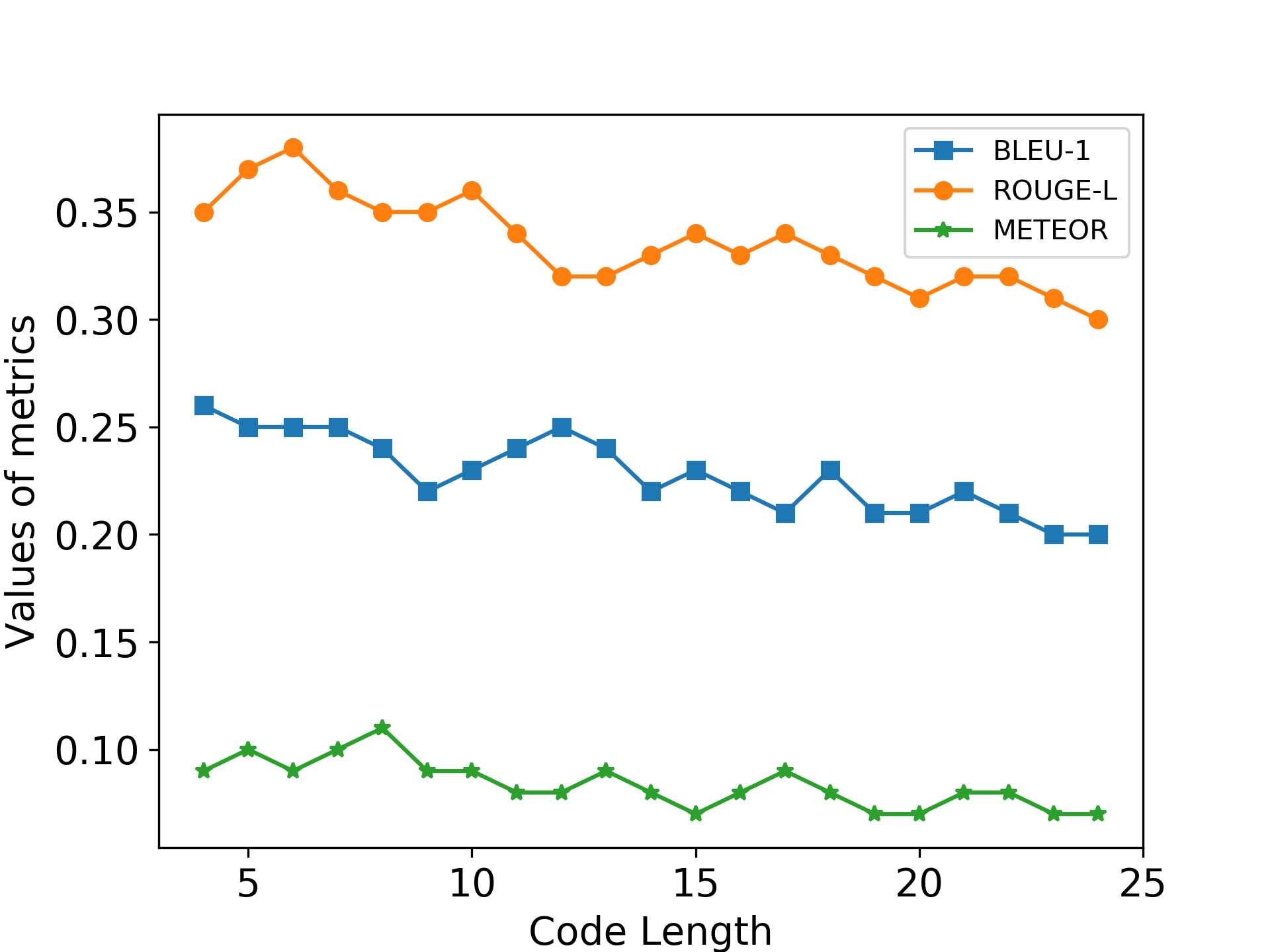}
        \caption{Performance of SolcTrans under different code lengths.}\label{fig:codeDP}
    \end{subfigure}
    \begin{subfigure}{0.45\linewidth}
        \includegraphics[width=\linewidth]{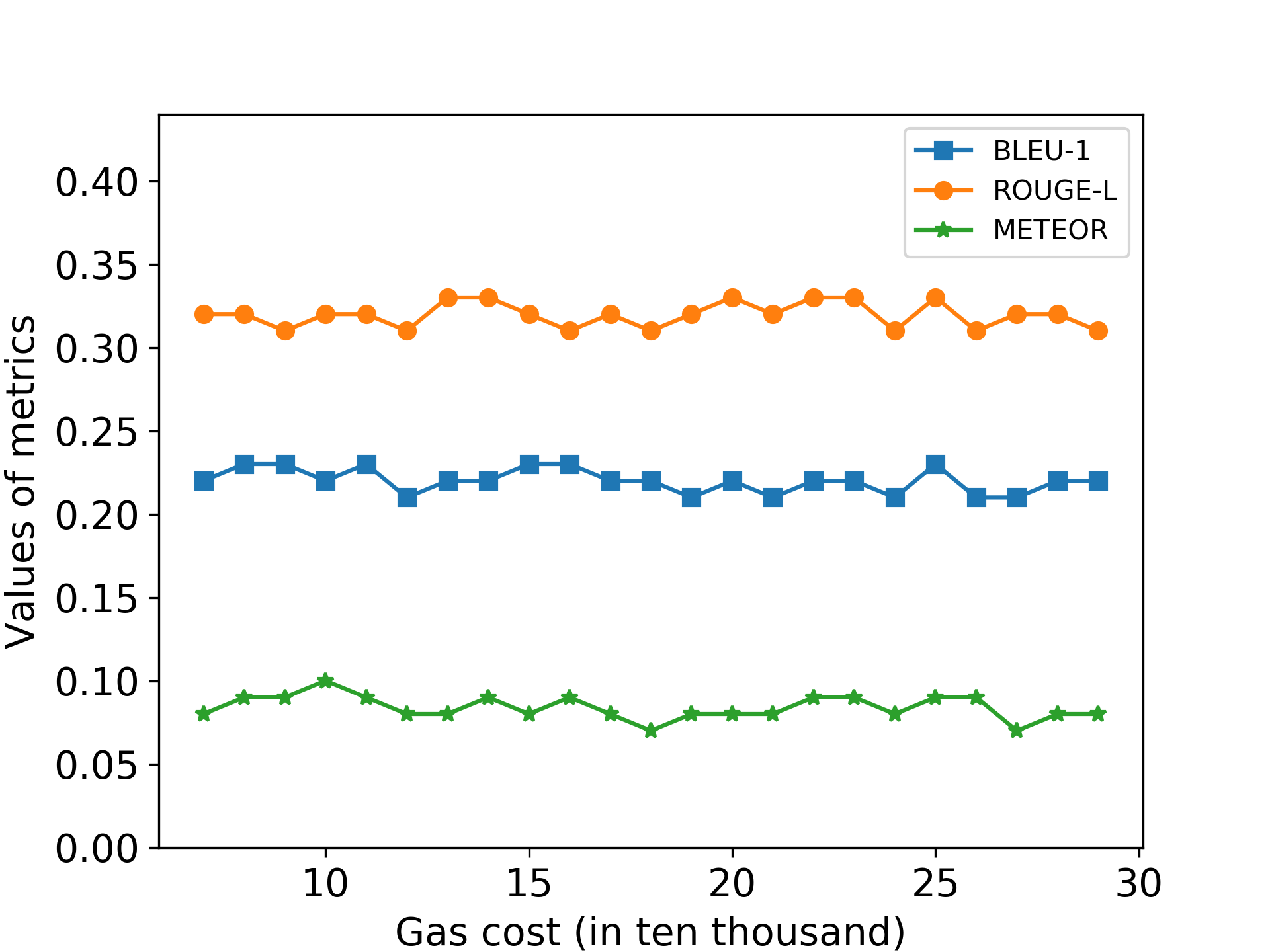}
        \caption{Performance of SolcTrans under different gas cost.}\label{fig:gasDP}
    \end{subfigure}
    \caption{Performance of SolcTrans under different code lengths and gas cost.}\label{fig:trend}
\end{figure*}

Since there is no parallel corpus large enough to train state-of-the-art neural machine translation models that can be used for Solidity code translation tasks, comparative experiments are yet available. To answer Q1 more intuitively, we conducted a qualitative analysis on some representative outputs of SolcTrans.

Table~\ref{tab:examples} shows some examples of translation outputs. The qualitative analysis reveals that SolcTrans can accurately translate common expressions such as assignments and function calls. For special elements, such as functions and variables, which always exist in the global namespace, the quality of translation depends on the preset templates. SolcTrans can also capture the structure information of \emph{ifstatement} and \emph{for loop} and generate correct and coherent descriptions. Sometimes the structure and word usage of the generated sentence are very different from those of the reference sentence, but they both describe the code snippet correctly, as shown in the third and fourth examples. The reason is that developers have different writing styles, while the translation outputs of SolcTrans are always consistent. 

To answer Q2, we further evaluated the performance of SolcTrans under different lengths of code snippets and gas cost of functions. Fig.~\ref{fig:trend} shows the trend of BLEU-1, ROUGE-L, and METEOR with the code length and gas cost. According to Fig.~\ref{fig:codeDP}, the translation quality reflected by the three metrics exhibits a slight downward tendency as the code length increases, because long sentences generated from CFG may contain more omissions and ambiguities. Another reason is that longer code snippets refer to more AST nodes, resulting in a higher probability of parsing error. However, the loss of translation quality under different code lengths did not exceed 20\% compared with the highest point. This result means that the translation quality of SolcTrans remains relatively consistent even as the code length increases. According to Fig.~\ref{fig:gasDP}, the translation quality barely changes with gas cost. The reason is that the gas cost can only indicates the number of operation steps, rather than the complexity of the code. For example, a function with nested loops has high gas cost, while loop structures are easy to parse. In addition, for any smart contract less than 3,200 lines (the length of the longest smart contract we have collected) and function less than 300,000 units of gas cost, the processing time of SolcTrans is less than 250 milliseconds.

\section{User study}
We carried out a user study among programmers and non-programmers to evaluate the usefulness of SolcTrans. In the following, we describe the survey design before showing the evaluation results.
\subsection{Survey Design}
The survey was designed in the form of a questionnaire. We performed stratified sampling from our data set based on the code length and selected a total of 10 contract snippets for evaluation. For each contract snippet, we presented the produced low-level code descriptions generated by SolcTrans as in-line comments. According to the standards for assessing source code comments~\cite{DBLP:conf/iwpc/SteidlHJ13, DBLP:journals/tse/McBurneyM16}, the survey focuses on three metrics: Accuracy, Readability, and Intelligibility. Our questions correspond to these three metrics, respectively, as Table~\ref{tab:questions} shows. The optional answer for each question is 1 to 5 (1 - Strongly Disagree, 2 - Disagree, 3 - Borderline, 4 - Agree, 5 - Strongly Agree). In addition to the rating, volunteers can provide supplementary explanations to their answers. 

\begin{table}[h]
\caption{The questions in the questionnaire.}
\label{tab:questions}
\begin{tabular}{c|l}
\hline
Metric          & \multicolumn{1}{c}{Question} \\ \hline
Accuracy        & The comments are accurate and concise.       \\
Readability     & The comments are easy to read.        \\
Intelligibility & The comments help me understand the code.        \\\hline
\end{tabular}
\end{table}

We recruited 20 volunteers through social media to participate in our survey. Ten of them are programmers, while others have no programming background. Among the programmers, 6 of 10 have at least two years of programming experience in Solidity, 4 of 10 are novice developers. Volunteers were asked to read the code-description pairs and record their answers to questions as the overall evaluation results. Non-programmers were excluded from Accuracy related questions.

\subsection{Results Analysis}
Table~\ref{tab:survey} shows the rating results of the survey.

\textbf{Accuracy.} Among ten programmers, there were 8 of them who rated ``Agree (3)'' and ``Strongly Agree (5)''. The majority of the programmers agreed that the generated comments accurately describe the corresponding code lines. The most frequently mentioned issue is that the words derived from camel-case variables are sometimes inaccurate. For example, a statement \textsf{``biddingEnd = now + \_biddingTime''} was translated as ``bidding end is now added to bidding time''. The variable ``biddingEnd'' should be translated as ``the time of bidding end'' rather than ``bidding end'' here. Such confusions are the results of poor naming conventions.

\textbf{Readability.} More than half of the programmers (6) and non-programmers (7) rated ``Agree'' and ``Strongly Agree'' in the Readability of translation outputs. Almost all the participants thought that short sentences were easy to read and understand. Negative reviews mainly focused on long sentences with more than 15 words, which is in line with the trend of Figure~\ref{fig:codeDP}. The generated sentences are generally considered complicated for the source code with a multi-layer nested structure. 

\textbf{Instructiveness.} Although most of the volunteers affirmed the Intelligibility of generated comments, non-programmers rated more positively (4 ``Agree'' and 6 ``Strongly Agree'') than programmers (5 ``Agree'' and 2 ``Strongly Agree''). Two experienced programmers rated Disagree. The generated comments are less useful to experienced programmers because they can read smart contracts without the help of such low-level descriptions. Novice developers rated positively since the generated comments explained system functions and keywords they were unfamiliar with. For non-programmers, generated comments significantly helped them understand the purpose and logic of the code snippets.

\begin{table*}[]
\caption{Statistical results of ratings.}
\label{tab:survey}
\begin{tabular}{l|c|ccccc}
\hline
                                 & Metric          & Strongly Disagree & Disagree & Borderline & Agree & Strongly Agree \\ \hline
\multirow{3}{*}{Programmers}     & Accuracy        & 0                 & 0        & 2          & 3     & 5              \\ \cline{2-7} 
                                 & Readability     & 0                 & 1        & 3          & 4     & 2              \\ \cline{2-7} 
                                 & Intelligibility & 0                 & 2        & 1          & 5     & 2              \\ \hline
\multirow{2}{*}{Non-programmers} & Readablility    & 0                 & 1        & 2          & 5     & 2              \\ \cline{2-7} 
                                 & Intelligibility & 0                 & 0        & 0          & 4     & 6              \\ \hline
\end{tabular}
\end{table*}

\section{Conclusion and future work}
This paper proposes SolcTrans, an approach that can automatically translate Solidity code into natural languages. Different from deep learning methods used for generating summarization of code, SolcTrans focuses on producing low-level code descriptions which can be used as in-line comments, aiming to help people without programming backgrounds understand the logic details of smart contracts. SolcTrans traverses the AST to translate code elements into corresponding words or phrases and aggregates them into readable sentences through CFG. The experimental and user study results show that the comments generated by SolcTrans are accurate and helpful. Our manuscript creates a paradigm for future studies of the machine translation of smart contracts. 

We identified several directions for further work. First, we will continue to expand the Solidity source code corpus with parallel reference sentences. When the corpus becomes large enough, it will be available to train state-of-the-art deep learning-based machine translation models. Then, we will conduct comparative experiments to show the performance of SolcTrans more comprehensively. Second, due to the complexity of Solidity, the code statements covered by the current CFG are not extensive enough. For example, we do not handle exception handling statements like \textsf{try\ldots catch\ldots} and some system functions included in AST. We plan to continuously improve the CFG to support more significant AST paths and nodes. We will also consider integrating grammar proofing tools to optimize the generated sentences in future works.

\bibliographystyle{ACM-Reference-Format}

\bibliography{cas-refs}


  \onecolumn
\appendix

\lstset{linewidth=6cm}
\begin{table}
\section{Common NodeTypes and Core Attributes}
  \renewcommand\arraystretch{1}
  \caption{Common NodeTypes and Core Attributes (with the code snippet in Listing 1 taken as an example).}
  \label{tab:nodeTypes}
  \begin{tabular}{cll}
    \toprule
    nodeType & \makecell[c]{Corresponding Code Element in Listing 1} &  \makecell[c]{Core Attributes}\\
    \midrule
    FunctionDefinition &  \begin{lstlisting}[language=Solidity,numbers=none]
function withdraw() public {
    uint amount = deposits[msg.sender];
    if (amount > 0) {
        deposits[msg.sender] = 0;
        msg.sender.transfer(amount);
    }
}
\end{lstlisting} & \begin{tabular}[c]{@{}l@{}}- name\\ - body (the inner content of the function)\\ - isConstructor (Boolean value)\\ - parameters\\ - returnParameters\end{tabular}\\\specialrule{0em}{3pt}{3pt}\hline
    Block & \begin{lstlisting}[language=Solidity,numbers=none]
uint amount = deposits[msg.sender];
if (amount > 0) {
    deposits[msg.sender] = 0;
    msg.sender.transfer(amount);
}
\end{lstlisting} &  \begin{tabular}[c]{@{}l@{}} - statements (all the expressions in the function)\end{tabular}\\\specialrule{0em}{1.5pt}{1.5pt}\hline
    VariableDeclarationStatement & \begin{lstlisting}[language=Solidity,numbers=none]
uint amount = deposits[msg.sender]
\end{lstlisting} &  \begin{tabular}[c]{@{}l@{}} - declarations \\- initialValue\end{tabular}\\\hline
    VariableDeclaration & \begin{lstlisting}[language=Solidity,numbers=none]
uint amount
\end{lstlisting} &  \begin{tabular}[c]{@{}l@{}} - name \\- typeDescription\end{tabular}\\\hline
    IndexAccess & \begin{lstlisting}[language=Solidity,numbers=none]
deposits[msg.sender]
\end{lstlisting} &  \begin{tabular}[c]{@{}l@{}} - baseExpression \\- indexExpression\end{tabular}\\\hline
    MemberAccess & \begin{lstlisting}[language=Solidity,numbers=none]
msg.sender
\end{lstlisting} &  \begin{tabular}[c]{@{}l@{}} - expression \\- memberName\end{tabular}\\\hline
    Identifier & \begin{lstlisting}[language=Solidity,numbers=none]
msg
\end{lstlisting} &  \begin{tabular}[c]{@{}l@{}} - name\end{tabular}\\\specialrule{0em}{1pt}{1pt}\hline
    IfStatement & \begin{lstlisting}[language=Solidity,numbers=none]
if (amount > 0) {
    deposits[msg.sender] = 0;
    msg.sender.transfer(amount);
}
\end{lstlisting} &  \begin{tabular}[c]{@{}l@{}} - condition \\- falseBody \\- trueBody\end{tabular}\\\specialrule{0em}{1pt}{1pt}\hline
    BinaryOperation & \begin{lstlisting}[language=Solidity,numbers=none]
amount > 0
\end{lstlisting} &  \begin{tabular}[c]{@{}l@{}} - operator \\- leftExpression \\- rightExpression\end{tabular}\\\hline
    Literal & \begin{lstlisting}[language=Solidity,numbers=none]
0
\end{lstlisting} &  \begin{tabular}[c]{@{}l@{}} - value\end{tabular}\\\specialrule{0em}{1pt}{1pt}\hline
    ExpressionStatement & \begin{lstlisting}[language=Solidity,numbers=none]
deposits[msg.sender] = 0;
msg.sender.transfer(amount)
\end{lstlisting} &  \begin{tabular}[c]{@{}l@{}} - expression\end{tabular}\\\specialrule{0em}{1pt}{1pt}\hline
    Assignment & \begin{lstlisting}[language=Solidity,numbers=none]
deposits[msg.sender] = 0
\end{lstlisting} &  \begin{tabular}[c]{@{}l@{}} - operator\\- leftHandSide\\- rightHandSide\end{tabular}\\\hline
    FunctionCall& \begin{lstlisting}[language=Solidity,numbers=none]
msg.sender.transfer(amount)
\end{lstlisting} &  \begin{tabular}[c]{@{}l@{}} - argumentTypes\\- expression\end{tabular}\\

    \bottomrule
  \end{tabular}
\end{table}

\lstset{linewidth=7cm}
\begin{table*}
\section{Examples of special statements translation templates.}
  \renewcommand\arraystretch{1}
  \caption{Examples of special statements translation templates}
  \label{tab:sps}
  \begin{tabular}{l @{}p{8cm}<{\raggedright}}
    \toprule
    \makecell[c]{Special Statement} & \makecell[c]{Translation Template}\\
    \midrule
    
    \begin{lstlisting}[language=Solidity,numbers=none]
blockhash(uint blockNumber) returns (bytes32)
\end{lstlisting} & the hash of the block \emph{blockNumber}\\\hline

    \begin{lstlisting}[language=Solidity,numbers=none]
block.coinbase (address payable)
\end{lstlisting} & the address of the current block miner\\\hline

    \begin{lstlisting}[language=Solidity,numbers=none]
block.difficulty (uint)
\end{lstlisting} & the current difficulty for mining a block \\\hline

    \begin{lstlisting}[language=Solidity,numbers=none]
block.gaslimit (uint)
\end{lstlisting} & the limit of the total gas usage in the current block \\\hline
    \begin{lstlisting}[language=Solidity,numbers=none]
block.timestamp (uint)
\end{lstlisting} & current timestamp of the block, represented as seconds since unix epoch\\\hline
    \begin{lstlisting}[language=Solidity,numbers=none]
msg.sender
\end{lstlisting} & user\\\hline
    \begin{lstlisting}[language=Solidity,numbers=none]
msg.value
\end{lstlisting} & weis sent by the user\\\hline
    \begin{lstlisting}[language=Solidity,numbers=none]
tx.gasprice(uint)
\end{lstlisting} & the single gas unit’s price set by the creator of the transaction\\\hline
    \begin{lstlisting}[language=Solidity,numbers=none]
tx.origin(address)
\end{lstlisting} & the address of the original external account that started the transaction\\\hline
    \begin{lstlisting}[language=Solidity,numbers=none]
gasleft() returns (uint256)
\end{lstlisting} & the remaining gas of the current call\\\hline
    \begin{lstlisting}[language=Solidity,numbers=none]
bytes.concat(...) returns (bytes memory)
\end{lstlisting} & concatenate \ldots~to one byte array\\\hline
    \begin{lstlisting}[language=Solidity,numbers=none]
assert(bool condition)
\end{lstlisting} & abort execution and revert state changes if \emph{condition} is not met\\\hline
    \begin{lstlisting}[language=Solidity,numbers=none]
require(bool condition, string memory message)
\end{lstlisting} & revert if \emph{condition} is not met, then provides \emph{memory message}\\\hline
    \begin{lstlisting}[language=Solidity,numbers=none]
addmod(uint x, uint y, uint k) returns (uint)
\end{lstlisting} & compute $(x + y)~\%~k$ \\\hline
    \begin{lstlisting}[language=Solidity,numbers=none]
mulmod(uint x, uint y, uint k) returns (uint)
\end{lstlisting} & compute $(x * y)~\%~k$ \\\hline
    \begin{lstlisting}[language=Solidity,numbers=none]
ripemd160(bytes memory) returns (bytes20)
\end{lstlisting} & compute RIPEMD-160 hash of the \emph{memory}\\\hline
    \begin{lstlisting}[language=Solidity,numbers=none]
abi.encode(...) returns (bytes memory)
\end{lstlisting} & \ldots~encoded by abi\\\hline
    \begin{lstlisting}[language=Solidity,numbers=none]
abi.decode(bytes memory encodedData) returns (...):
\end{lstlisting} & the \emph{encodedData} in \ldots~forms decoded by abi\\ 
    \bottomrule
  \end{tabular}
\end{table*}
\end{document}